\documentclass{icrc}

\usepackage{times}
\usepackage{graphicx} % when using Latex and dvips
\begin{document}

\title{TeV Blazars and Cosmic Infrared Background Radiation}
\author[1]{F.A. Aharonian}
\affil[1]{Max-Planck-Institut f\"ur Kernphysik, Heidelberg}

\correspondence{felix.aharonian@mpi-hd.mpg.de}

\firstpage{1}
\pubyear{2001}

% \titleheight{11cm} % uncomment and adjust in case your title block
                     % does not fit into the default and minimum 7.5 cm

\maketitle

\begin{abstract}
The recent developments in  studies of TeV radiation from  {\em blazars} 
are highlighted  and the implications of these results for derivation of 
cosmologically important information about the {\em cosmic  
infrared background radiation} are discussed.    
\end{abstract}

\section{Introduction}

Since early 90's  many papers have been published with an unusual 
combination of   two keywords -  {\em Blazars} and 
{\em Cosmic-Infrared-Background} (CIB)  radiation.  
Formally,  one may argue that there is no apparent link between these 
two topics.      

\noindent
$\bullet$ The blazars constitute a  sub-class of AGN
dominated by highly variable (several hours or less)  
components of broadband  (from radio to gamma-rays) 
{\sl non-thermal  emission}  produced in  relativistic jets 
pointing close to the line of sight.   

\noindent
$\bullet$ CIB  is a part of the overall diffuse extragalactic 
background radiation (DEBRA)    
dominated by {\sl thermal emission}  components produced by stars and dust,
and accumulated over the entire history of the Universe.  

\vspace{1mm}

While the blazars may serve  as ideal laboratories for  study 
of  MHD structures and  particle acceleration processes in
relativistic jets,   CIB carries  crucial cosmological  information 
about the  formation epochs and history of evolution of galaxies. 
To a large extent,  these two topics  are  relevant to 
quite independent  areas  of modern  astrophysics and cosmology.    
Yet,  the current studies of CIB and blazars, more specifically 
the sub-population of blazars emitting TeV gamma-rays ({\em TeV blazars}),   
are tightly coupled  through the intergalactic (IG) 
absorption  of TeV radiation by infrared photons of DEBRA. 

The  astrophysical/cosmological  importance of  this interesting effect 
(Nikishov, 1962; Gould and Schreder, 1966; Jelly, 1966; Stecker et al., 1992) 
was  clearly recognized  after the discovery of TeV  
$\gamma$-rays from two BL Lac objects -- Mkn~421 and Mkn~501 
(for review see e.g. Vassiliev 2000).  

\section{Cosmic Infrared Background Radiation}

CIB  basically consists of two emission components 
produced by stars and partly absorbed/re-emitted by  dust during 
the entire history of evolution of galaxies. Consequently,  two  distinct bumps  
in the spectral energy distribution (SED)  of red-shifted radiation at 
near infrared (NIR)  $\lambda \sim$ 1-2  $\mu \rm  m$ and    far infrared (FIR) 
$\lambda \sim$ 100-200 $\mu \rm m$ wavelengths, and a mid infrared (MIR) 
``valley'' between these bumps are expected. 
Because of the heavy contamination  caused by foregrounds of different origin,
predominantly  by the  zodiacal (interplanetary dust) light,  
the measurements of CIB contain large uncertainties.  
Moreover,   these results  only conditionally can be treated  as 
{\it direct measurements},  because  their interpretation  primarily   
depends on the  modeling and removal of these foregrounds.
Therefore,  the direct  observations  of CIB generally 
allow derivation  of the 
flux {\sl upper limits} 
rather than detection  of positive residual signals.  In Fig.~1 we  show  
the  CIB fluxes based on the latest reports, and refer the reader to  the  
review article by  Hauser and Dwek (2001) on 
the  current status of  direct observations of CIB.

Presently,  the most  reliable  results  are obtained from  the COBE 
observations at NIR and FIR 
domains where the contribution of  the zodiacal light  becomes 
comparable with the CIB flux. 
This  concerns, first of all, the FIR wavelengths at 
140 and  240  $\mu \rm m$
(Hauser et al., 1998, Schlegel et al., 1998, Lagache et al., 1999),  
and perhaps also at 100 $\mu \rm m$ (Lagache et al., 1999; 
Finkbeiner et al., 2000)
where the results of 3 different groups are in a reasonable  
agreement with each other,   but   somewhat higher 
compared to the theoretical  predictions.  If the reference  
of these fluxes to the truly 
extragalactic background  is correct,  this would  imply that most of the 
star formation in the early Universe  must occur in  highly obscured, 
dusty environments.   

%
%  ------------- Fig.1: CIB -------------
\begin{figure}[tb]
\begin{center}
\includegraphics[width=0.9\linewidth]{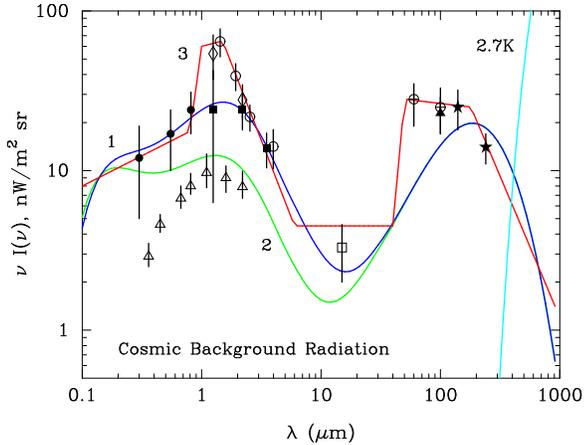}
\caption{SED of Cosmic Background Radiation.
The reported fluxes at 140 and 240 $\mu \rm m$ (stars) are from 
Hauser et al. (1998), at 100 $\mu \rm m$ (filled triangle) from
Lagache et al. (1999), at   60 and 100 $\mu \rm m$ 
(open circles) from  Finkbeiner et al. (2000).
The point  at 15 $\mu \rm m$ (open square) is derived from
the ISOCAM source counts  (Franceschini et al., 2001).
The fluxes reported at NIR and optical wavelengths are 
shown by filled squares (Wright and Johnson, 2001), 
open circles (Matsumoto, 2000),  open diamonds  (Cambresy et al., 2001),
and filled circles (Bernstein, 1998). The open triangles from   Pozzetti et  al. (1998)
correspond to lower limits. While the reference models 2 and 3 
should be considered as lower and upper limits, the curve 1 
may  be treated as the ``model of choice''.} 
\end{center}
\end{figure}

Detections of CIB are claimed also at NIR - at 2.2  $\mu \rm m$ and   
3.5 $\mu \rm m$ wavelengths
(Dwek and Arendt, 1998; Gorjian et al., 2000; 
Wright and Johnson,  2001).  
These fluxes  together with the J-band  upper limit at $1.25 \ \mu \rm m$
(Wright and Johnson,  2001) and the fluxes of the optical light
derived from the HST data  below $0.8  \ \mu \rm m$ (Bernstein, 1998),
agree with the recent theoretical calculations by  Primack et al.  (2001).
An  independent analysis of  the COBE data  by Cambresy et al. (2001), as well as  
the results of the Japanese IRTS satellite (Matsumoto et al. 2001) are in agreement 
with the fluxes  at   $2.2 \ \mu \rm m$   and $3.5 \ \mu \rm m$ reported 
by  Wright and Johnson  (2001), but at  shorter wavelengths the claimed fluxes 
are  noticeably higher compared to other measurements (see Fig.~1). 

The ``best guess'' estimate of the 
SED  at  optical/NIR wavelengths of about 20-50 $\rm nW/m^2 sr $ 
is comparable with the FIR flux of about 40-50 $\rm nW/m^2 sr $ 
(Madau and Pozzetti,  2000).  
This indicates that an essential part of 
the energy radiated by stars is absorbed and re-emitted by cold dust.  
Our current knowledge  of  MIR,   which carries information about the 
warm dust component,  is quite limited.  The only flux estimates   
in this band derived from the ISOCAM source counts 
at 6 and 15 $\rm \mu m$ (Franceschini et al., 2001)  apparently  
should be taken as lower limits. 
%
%  ------------- Fig.2: Mean free path -------------
\begin{figure}[tb]
\begin{center}
\includegraphics[width=0.7\linewidth]{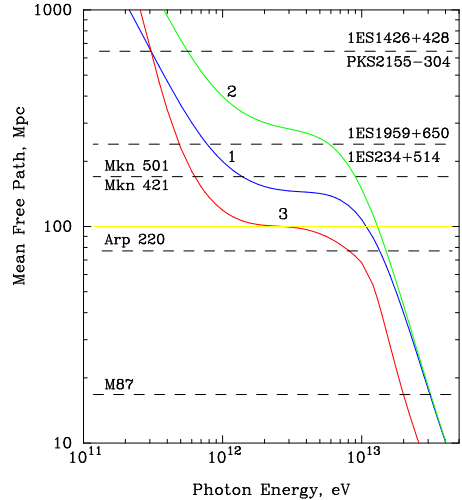}
\caption{Mean free-path of VHE  $\gamma$-rays
in the intergalactic medium calculated for 3 CIB models presented in Fig.~1.} 
\end{center}
\end{figure}

These measurements  stimulated 
new  calculations of CIB 
based on different  phenomenological  
and theoretical  (backward, forward, 
cosmic chemical evolution, semi-analytical, {\em etc.}) approaches. 
In spite of certain achievements,  the ability of  current 
(most successful) models to reproduce  the CIB  measurements 
nevertheless should not be overstated,  because all these models 
contain  a number of adjustable parameters and 
in fact are ``primarily designed for that purpose''   
(Hauser and Dwek, 2001). 

In Fig.~1 three  CIB models  are shown which 
{\sl provisionally} may be assigned  as    
``nominal'' (1),  ``low'' (2), and ``extreme''  (3) 
descriptions  of CIB.  The curve 2 corresponds to the 
so-called ``LCDM-Salpeter''  model  of Primack et al. (1999),  
but assuming  twice  larger contribution of the dust  component.  
Similar  NIR spectra 
with a flux of about  $10  \ \rm nW/m^2 sr$ at 1 $\mu \rm m$  have 
been assumed  by  De Jager and Stecker (2001) and Franceschini et al. (2001). 
The curve 1  is close to the prediction  by Primack et al. (2001) based on the 
so-called  ``Kennicut''  stellar initial mass function (IMF).    
And  finally,  the curve 3 is  ``designed''  to  match the extreme fluxes reported 
by  Matsumoto (2000) and   Cambresy et al.  (2001) at NIR
and Finkbeiner et al. (2000) at FIR. 

\section{Absorption of TeV gamma-rays in CIB}

To calculate the mean free path of $\gamma$-rays $\Lambda(E)$ in the IGM 
one must convolve the CIB photon number distribution  $n(\epsilon)$ with the 
pair production cross-section.   Because of the narrowness 
of the latter,  for broad-band photon spectra 
over half the interactions of a $\gamma$-ray photon of energy $E$ 
occur with a quite narrow interval of 
target photons   
$\Delta \lambda \sim (1 \pm 1/2)  \lambda^\ast$ centered on  
$\lambda^\ast \approx 1.5 (E/{\rm 1 \ TeV}) \ \rm \mu m$. 
This gives  a convenient approximation  for the 
optical depth   $\tau(E)=d/\Lambda(E)$ for $\gamma$-rays 
emitted by a source at a distance $d$: 
$$
\tau(E) \simeq 1   
\left(\frac{u_{\rm CIB}(\lambda^\ast)}{10 \ \rm nW/m^2 sr}\right)  
\left(\frac{E}{\rm 1 \ TeV}\right)  
\left(\frac{z}{0.1}\right)   H_{60}^{-1} \ , 
$$
where $u_{\rm CIB}=\nu I(\nu)=\epsilon^2 n(\epsilon)$
is the SED of CIB, $z$ is the source redshift,   
$H_{60}$ is the Hubble constant normalized to  $60 \ \rm km/s \ Mpc$.
No deviation of the observed $\gamma$-ray spectrum 
from  the intrinsic (source) spectrum, e.g. by a factor of $\leq 2$, would  imply
$\tau(E) \leq \ln 2$, and consequently provide an  upper limit 
on the CIB flux at $\lambda^ \ast (E)$. For example,  this condition 
for 1 TeV $\gamma$-rays  from   Mkn 421 or Mkn 501 
($z \simeq 0.03$) gives  $u_{\rm CIB} \leq 20 \ \rm nW/m^2 sr$ 
around 1-2 $\mu \rm m$.  
Since  this estimate is quite close to the recent NIR measurements (see Fig.~1), 
we may conclude  that already at 1 TeV  we detect  significantly 
absorbed $\gamma$-radiation  from Mkn 421 and Mkn 501. This is demonstrated   
in Fig.~2 where  we present  accurate numerical calculations of the mean free
path  of $\gamma$-rays  for three  CIB reference models shown 
in Fig.~1. The horizontal lines indicate
the distances to  the reported  TeV blazars (see below), as well as to 2    
nearby prominent extragalactic  objects -- the radiogalaxy M 87 and the 
ultraluminous starburst galaxy  Arp 220.
The curve 2 calculated for the CIB lower limit (model 2 in Fig.~2),
should be treated as an upper limit for the mean free path.
This implies that  we cannot ignore the IG-absorption of TeV 
$\gamma$-rays  from sources located beyond 100 Mpc.  On the other hand,
this gives  a unique chance to extract information about CIB by 
detecting absorption features in TeV  spectra of  extragalactic objects.
%
%  ------------- Fig.3: Corrected Mkn 501 -------------
\begin{figure}[tb]
\begin{center}
\includegraphics[width=0.7\linewidth]{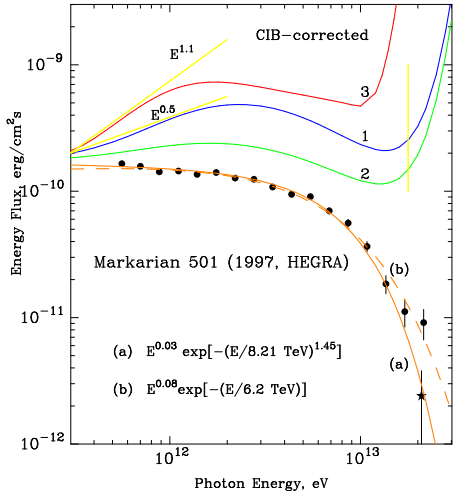}
\caption{Absorption-corrected SED  of  Mkn 501
reconstructed for the analytical presentation 
of the measured spectral points  given by equation (a) and assuming 
3 different CIB models  shown in Fig.~1.  
The experimental points (filled circles) 
correspond  to the 1997 time-averaged spectrum of Mkn 501 (Aharonian et al.,  1999a).
The star corresponds to the recently re-estimated flux  
at $E \approx 21 \ \rm TeV$ (Aharonian et al., 2001a). The vertical line at 17 TeV 
indicates the edge of the spectrum measured with high statistical 
significance.}
\end{center}
\end{figure}

Very often such a possibility is reduced  to the search 
for sharp cutoffs in the  energy spectra of extragalactic TeV sources 
(see e.g. Stecker et al., 1992).  However,   in many cases  this  
could be a  misleading  recommendation.  
Indeed,  the {\it cutoff} in a $\gamma$-ray spectrum
does not yet imply an  {\it evidence for the IG-absorption}
and vice versa, the {\it lack of the cutoff}  cannot be interpreted as  
{\it absence of IG-absorption}.  It is interesting to note that all 
CIB  models with conventional spectral shape between 1 and 10 
$\mu \rm m$, predict  almost  constant (energy-independent)  
mean free path of $\gamma$-rays, and correspondingly 
insignificant  spectral deformation   at energies  between 1 
and  several TeV (Primack et al.,  2001). 
The explanation of this effect   is straightforward. 
If the spectrum of background photons in a certain energy  interval has a 
power-law dependence,   $n(\epsilon) \propto \epsilon^{-\beta}$,  
the mean free path  in the corresponding $\gamma$-ray energy interval 
$\Lambda(E) \propto E^{1-\beta}$. 
Within $\rm 1 \mu m < \lambda < 10   \mu$ the CIB spectrum 
is  approximately  described by a power-law 
$n(\epsilon) \propto \epsilon^{-1}$
or $u_{\rm CIB} \propto \lambda^{-1}$ (see Fig.~1), therefore 
in the interval between 1 and several  TeV  
the $\gamma$-ray mean free path   only slightly depends on energy (Fig.~2).
In order to demonstrate the impact of the  
IG-absorption,  in Fig.~3 we show  the SED  
of Mkn 501 together with  the  absorption-corrected  
spectra   reconstructed for 3 CIB models  shown in Fig.~1.

\section{Observations of TeV Blazars}

Many nonthermal extragalactic objects representing different classes 
of AGN  and located  within 1  Gpc are  potential  
TeV sources.  First of all this concerns the BL Lac 
population of blazars of which two nearby representatives, 
Mkn 421  ($z=0.031$)  and Mkn 501 ($z=0.034$)  
are firmly established as TeV $\gamma$-ray emitters.
The current list of extragalactic TeV sources  contains 
4 more BL Lac objects with reported  signals at  4 to 6
sigma level:  1ES~2344+51 
[Whipple (Catanese et al., 1998)] at  $z=0.044$, 
PKS~2155-304 [Durham (Chadwick et al., 1999)] at $z=0.116$, 
1ES~1959+650 [Telescope Array  (Nishiyama et al., 1999) and 
HEGRA (G\"otting et al., 2001)] at $z=0.048$, and  
1ES~1426+428  [VERITAS (Horan et al., 2001) and HEGRA 
(G\"otting  et al., 2001)]  at $z=0.129$.  

Since the  discovery of TeV $\gamma$-radiation of  Mkn 421  
(Punch et al. 1992), this object has been subject
of intensive studies through multiwavelength observations.   
The TeV flux of the source is variable with typical average value
between $30 \%$ to $50 \%$ that of the Crab Nebula, but
with strong and rapid,  as short as 0.5 h,   flares 
(Gaidos et al., 1996)  which correlate  with the source
activity at other wavelengths.  Until the  
spectacular high state  of the source in 2001,  the spectral studies 
were based mainly on the data taken during
a quiescent or moderately high states. In particular, 
the spectrum measured by HEGRA during the 1998 ``ASCA'' campaign 
(Takahashi et al., 2000) is  fitted  by a steep power-law with photon 
index $\Gamma \simeq 3$ (Aharonian et al.,  1999b).  
The exceptionally bright and long-lasting activity 
of Mkn 421 in 2001 allowed the VERITAS  (Krennrich et al., 2001)
and HEGRA groups (Horns et al., 2001)
to  derive  the time-averaged gamma-ray spectrum of the source in the high state
which  up  to  $\sim 15$ TeV 
is described as  ${\rm d} N/{\rm d} E = K E^{\Gamma} \exp{(-E/E_0)}$,  
i.e. by the same  canonical  ``power-law with exponential cutoff'' function
found  earlier for the high-state, 
time-averaged spectrum of  Mkn 501 (Aharonian et al., 1999a).
The spectra of these  sources  are not, however,  identical.
In Fig.~4 we show the energy spectra  based on 
approximately 40,000 and 60,000 TeV $\gamma$-ray events 
detected by HEGRA during  the exceptionally high 
states of Mkn 501 in 1997 and Mkn 421 in 2001, respectively.
They are  described by different [$\Gamma, E_0$] combinations:  
$\Gamma=1.92 ~\pm 0.03 $ and $E_0=6.2 ~\pm 0.4  \, \rm TeV$
for Mkn 501 (Aharonian et al., 1999a) and   
$\Gamma=2.23 ~\pm 0.04 $ and $E_0=4.0. ~\pm 0.4  \, \rm TeV$
(Horns et al., 2001).  The power-law part of the spectrum of 
Mkn~421 is steeper,  and the exponential cutoff starts earlier.
Since both sources are located at approximately same distances, 
the difference in the cutoff energies may be interpreted  
as an evidence against  the hypothesis  which attributes the
cutoffs  to the  pure IG-absorption effect. 
%
%  ------------- Fig.~4 -----------Mkn 421/501
\begin{figure}[tb]
\begin{center}
\includegraphics[width=0.85\linewidth]{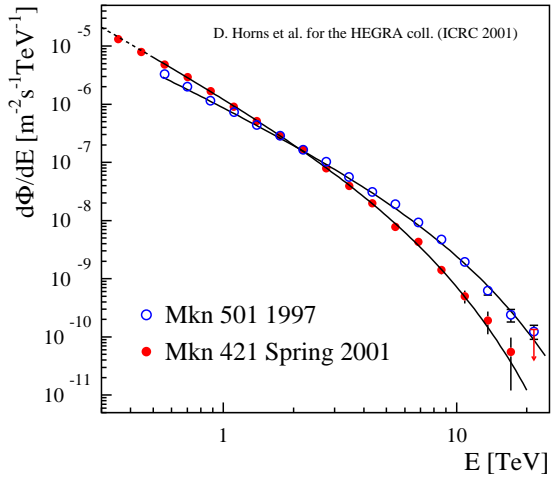}
\caption{Energy spectra of  Mkn 501 (Aharonian et al., 1999a) and 
Mkn 421 (Horns et al., 2001)  as measured by HEGRA CT system 
during the high states of sources in 1997 and 2001, respectively.}
\end{center}
\end{figure}

For the typical energy resolution of Cherenkov  telescopes of about 
$20 \%$  or  larger, one has to be careful with conclusions concerning 
the spectral shape  at energies beyond  $\approx 3 E_0$.  Obviously  for 
determination  of 
spectra with sharper  (super-exponential) cutoffs,  a significantly  better 
energy resolution is required.  Motivated by this,  HEGRA collaboration  recently 
re-analyzed  the ``1997 high-state''  spectrum of Mkn 501
using an improved method for energy reconstruction with 
$\simeq 12 \%$ resolution   achievable for  
the stereoscopic mode of observation  (Hofmann, 2000). 
The new analysis   confirms the result of 
the previous study up to  17 TeV,  
but constrains stronger the flux at 21 TeV (Aharonian et al., 2001a)
With this new point,   the Mkn 501 spectrum at very high energies is better described
by a super-exponential cutoff (compare  curves (a) and (b) in Fig.~3). 
%
%  ------------- Fig.5 -------------RXTE/WHipple
\begin{figure}[tb]
\begin{center}
\includegraphics[width=0.7\linewidth]{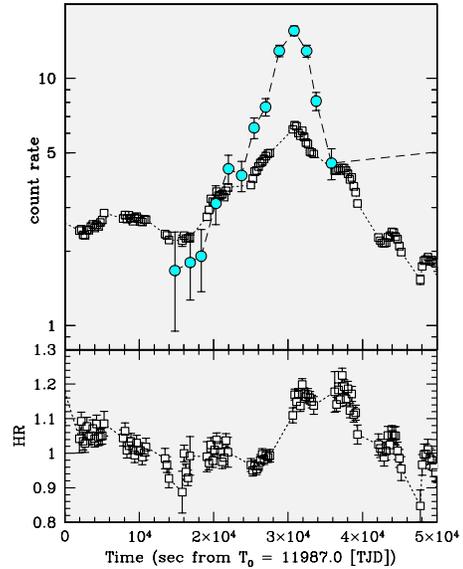}
\caption{Simultaneous X-ray and $\gamma$-ray observations of the 
2001 March 19 flare  by RXTE (squares) and VERITAS (circles). 
The bottom panel shows the RXTE/PCA
``6-11 keV/3.5-4.5 keV''  hardness ratio (from Fossati et al., 2001).}
\end{center}
\end{figure}
%

%  ------------- Fig.6 -------------
\begin{figure}[tb]
\begin{center}
\includegraphics[width=0.75\linewidth]{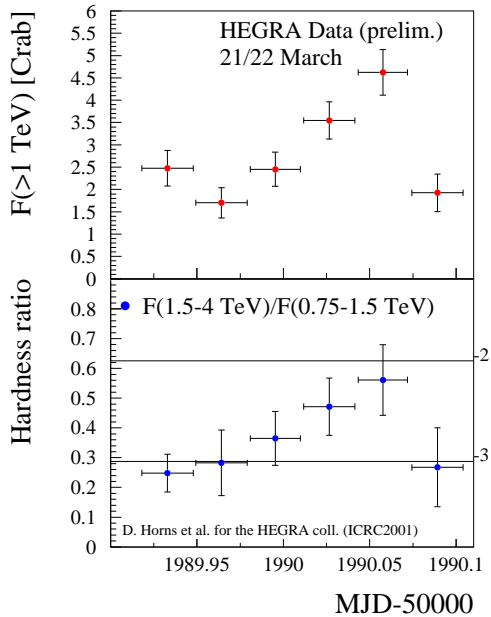}
\caption{Time-evolution of the absolute flux and the hardness
ratio of TeV radiation of Mkn 421 observed by the HEGRA IACT system 
on March 21/22, 2001 (Horns et al., 2001).}    
\end{center}
\end{figure}

Despite the huge  overall photon statistics,
the spectra of both sources  above 10 TeV can be determined   
using  data accumulated over long periods of observations.   
Obviously,  the time-averaged spectra of variable sources 
obtained in such way can be considered as {\sl astrophysically meaningful}
provided that the  spectral shape is essentially time-independent,
i.e. does not correlate with the absolute flux. 
Remarkably, it turned out that during the high state in 1997  
the shapes of daily spectra of Mkn 501remained essentially stable 
despite dramatic flux variations (Aharonian et al., 1999a), although   
for some specific flares significant 
spectral variations  cannot be excluded. 
For example, the CAT group    (Djannati-Atai  et al., 1999)
found  an evidence for a non-negligible 
hardness-intensity correlation  for the 1997 April 7 and 16  flares.

The  sensitivity of current ground-based detectors is not sufficient to
study  the  spectral variability of TeV radiation of 
Mkn 421 and Mkn 501 in  quiescent states  on timescales less than 
several days.  However,  even the time-averaged spectra
of these sources in low states are of certain interest.  
The HEGRA observations of Mkn 501 
in 1998 and 1999, when the source was at a $\approx 10$ 
times lower flux level compared to the average flux in 1997, 
revealed a noticeable steepening  (by $0.44 \pm 0.1$ in photon index) 
of the energy spectrum (Aharonian et al. 2001b).  This  seems to be true also
for Mkn 421, al least at low energies.
While the spectrum of this source in a  quiescent state has a steep power-law spectrum with 
photon index $\Gamma \simeq 3$,  the time averaged spectrum of 
the 2001 outburst is significantly flatter,  with $\Gamma \simeq  2.2$   
around 1 TeV.

The SEDs of BL Lacs  are  expected  to be 
very hard up to TeV  energies,  therefore the  imaging atmospheric 
Cherenkov telescopes (IACTs)
are nicely  suited to search for short  
signals from these objects. In particular,  the HEGRA, CAT 
and  VERITAS   IACTs  can follow  $\approx 10^{-11} \ \rm erg/cm^2 s$ 
(at 1 TeV)   flares of Mkn 421 and Mkn 501 
on timescales less than several hours, and thus are  
well-matched to the sensitivity and spectral coverage
of X-ray satellites like RXTE,  BeppoSAX and XMM 
for multiwavelength monitoring  of flux variations.
This is a key condition which makes the 
simultaneous X- and $\gamma$-ray observations 
meaningful and very important (e.g. Pian et al., 1998,  
Sambruna et al., 2000;  Maraschi et al., 1999; Kataoka et al., 1999; 
Krawczynski et al., 2000;  Takahashi et al., 2000).  
The best so far results in  this regard 
became available recently, after  the well coordinated multiwavelength 
campaigns of spectacular flares of Mkn 421 in 2001.
On several occasions,  truly simultaneous observations 
by   RXTE and TeV instruments with  durations up to 6 hours  per night 
were carried out.  A nice sample  of such events, the  2001 March 22 flare detected  
by RXTE and VERITAS  (Fossati et al., 2001) 
shows  a clear  keV/TeV correlation on sub-hour timescales (see Fig.~5).

For deep understanding of acceleration and radiation
processes in jets,  it is crucial  to search for {\em spectral} variability on 
timescales  comparable to  the characteristic dynamical  times of about   
1~h.  The HEGRA IACT system is able  to perform such studies,  
if  the flux at  1 TeV  exceeds  1-2 Crab units, 
$J_{\rm Crab}(E \geq 1 \ \rm TeV)  \simeq 1.7 \times 10^{-11} \, \rm ph/cm^2 s$.
The strongest flares of Mkn 421 in 2001  provide us with 
such  unique data.  
The preliminary spectral analysis of observations of these 
flares (Horns et al., 2001)  shows   a complex picture. 
If  the spectra  of some of these 
flares do not show noticeable spectral  changes, the others, in 
particular the  21/22 March 2001 flare    
demonstrates an impressive correlation  of the hardness ratio 
with the  absolute flux (see Fig.~6).

\section{Radiation mechanisms of TeV gamma-rays}

%
%  ------------- Fig.7 -------------
\begin{figure}[tb]
\begin{center}
\includegraphics[width=0.75\linewidth]{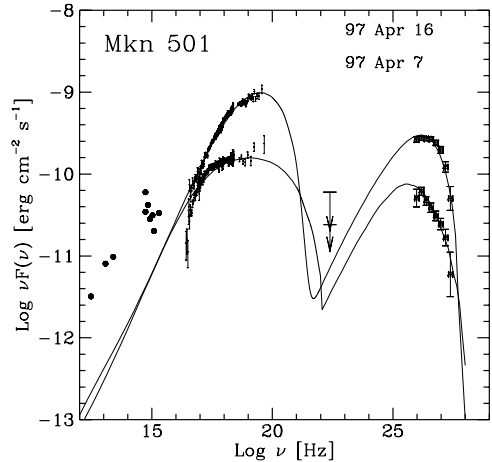}
\caption{Interpretation of the quasi-simultaneous X-ray 
({\em BeppoSAX}) and TeV (CAT)  
observation  of 1997 April 7 and 16 flares of 
Mkn 501 within the homogeneous SSC model (Tavecchio et al., 2001).
}
\end{center}
\end{figure}

The  studies of spectral evolution of TeV $\gamma$-rays and their 
correlations  with X-rays on sub-hour timescales  not only imply   
a new standard in  $\gamma$-ray astronomy,  but also indicate that   
the quality of the TeV data is approaching to the level 
which would allow us to follow and resolve simultaneously  
the predicted fluctuations in the putative synchrotron and IC emission 
components on timescales close to the shortest ones likely in these objects.  
This may have two important consequences (Coppi and Aharonian, 1999a): 

\noindent
$\bullet$~ Matching the observed X-ray/TeV light curves (as opposed 
to simply fitting snapshot spectra obtained many hours and days apart)
should provide a very stringent test of the so-called synchrotron-self Compton (SSC) 
model since we have two detailed handles on the {\it single electron} distribution
responsible for both emission components.
This test can rule out alternative  {\em hadronic}  models which are less attractive
but still viable options for explanation of TeV emission. 

\noindent
$\bullet$~If the SSC model works, it will be possible to fix 
the key model parameters and calculate the 
blazar's intrinsic spectrum, comparing the observed 
variations of {\sl absolute fluxes} and {\sl spectral shapes} with the predictions 
of self-consistent, time-dependent numerical codes.  This is a crucial point because  
with an estimate of the intrinsic TeV spectrum, then, and {\sl only then},  we can 
attempt  to  estimate the IG-absorption effect by comparing the intrinsic and 
observed spectra, and thus to get information about CIB.    

The observed  X/TeV correlations are often 
interpreted as  a strong argument in favor of SSC model, 
the simplified  one-zone version of which 
generally gives satisfactory fits to the observed 
X-ray and TeV spectra of Mkn 421 and Mkn 501 
(e.g. Mastichiadis and  Kirk, 1997; Bednarek and  Protheroe, 1999;
Kataoka et al., 1999;  Krawczynski et al., 2000; Tavecchio et al., 2001, {\em etc.}).
An example of  very  good spectral fits  for  the  1997  April  
flares of Mkn 501,  obtained for  fixed ``standard'' model parameters
$\delta_{\rm j}=10$ and $B \simeq 0.3 \ \rm G$ and changing only 
the maximum energy  of accelerated electrons,  is shown in Fig.~7 (Tavecchio et al., 2001).   
It should be noticed, however,  that   
the calculation in  Fig.~7 do not take into account the IG-absorption
of $\gamma$-rays which otherwise results in rather flat  intrinsic 
$\gamma$-ray spectra  (Coppi and Aharonian, 1999b; 
Konopelko et al., 1999; Guy et al, 2000).
In particular, any CIB model that predicts 
NIR fluxes close or higher  the CIB reference model no.1  
would require  unusually hard  intrinsic TeV spectrum (Guy et al., 2000) 
with photon index $\Gamma \leq 1.5$ (see Fig. 3).  Such spectra 
cannot be  easily   explained by the standard one-zone SSC model without 
invoking  extreme jet parameters   like Doppler factors  
$\delta_{\rm j} \sim 100$,  and very small B-field  $B \leq 0.01 \ \rm G$,  
which however  reduce  the radiation efficiency  to an 
almost unacceptably low level (Krawczynski   et al., 2001).
The situation is even more difficult   
for Mkn 421,  the X-ray  synchrotron cutoff of which never  extends  
beyond 10-20  keV.   This  significantly limits the freedom for 
assuming parameters  which would  allow hard  multi-TeV  IC spectra.
  
Despite these difficulties,  the leptonic  models (perhaps in their more 
sophisticated versions including the multi-zone SSC and/or External Compton
components) remain the generally preferred  concept for TeV blazars. 
These  models have two  attractive features:  {\em (i)} 
the capability of the  (relatively) well understood 
shocks to accelerate electrons to TeV energies (Sikora 
and Madejski, 2001; Pelletier, 2001)  and {\em (ii)} effective production of tightly 
correlated X-ray  and TeV  $\gamma$-ray emission components   
via  the synchrotron 
and inverse Compton radiation channels (e.g. Ulrich et al., 1997). 
 
The so-called hadronic models are generally lacking in  these virtues.
These models  assume that the observed $\gamma$-ray emission
is initiated by accelerated protons interacting with ambient mater  
(Bednarek, 1993; Dar and Laor, 1997; Pohl and Schlickeser, 2000),  
photon fields (Mannheim, 1996),  
magnetic fields  (Aharonian, 2000) or both (M\"ucke and Protheroe, 2001).  
The models involving interactions of protons with photon and B-fields   
require particle acceleration to extreme 
energies  exceeding $10^{19} \, \rm eV$ which  is possible if the  
acceleration time is  close to 
$t_{\rm acc}=\eta (r_{\rm g}/c)$ with $\eta \sim 1$ ($r_{\rm g}$ and $\eta$
are the so-called gyro-radius, and gyro-factor, respectively, $c$ is the speed of light).   
This  corresponds (independent of specific acceleration  
mechanism) to the maximum (theoretically possible) 
acceleration rates  which hardly can be achieved  by  the  
conventional diffusive shock acceleration mechanism 
(e.g. Pelletier, 2001, Sikora  and Madejski, 2001).
On the other hand,  the undisputed (for any TeV blazar model)  condition  of  high 
(close to $100 \%$)  efficiency  of  radiative cooling  of accelerated particles 
requires extreme parameters characterizing the  sub-parsec jets and their 
environments, in particular  very high (often, almost  unrealistic)
densities of the thermal plasma,  radiation and/or B-fields.  
But of course, these problems  should not prevent us from considering 
the hadronic models as  a viable  alternative to the leptonic models. 

It should be noticed in this regard that the very 
fact of strong X/TeV correlations do not yet exclude the 
hadronic  models, because the {\em secondary}  electrons -- the products
of interactions of accelerated protons and primary $\gamma$-rays  -- 
can be readily  responsible for the synchrotron X-radiation. 
Also, some of the hadronic models,
for example the   {\em synchrotron-proton} model 
can provide not only effective radiative (synchrotron) cooling but also  
quite good  TeV spectral fits.  
The proton-synchrotron model 
allows very hard intrinsic TeV  spectra  
and can rather naturally explain 
the stable spectral shape of TeV radiation  of Mkn 501 
during strong flares by the self-regulated synchrotron cutoff at 
$E_{\rm cut} \simeq 0.3  \eta^{-1} \delta_{\rm j}$ (Aharonian, 2000).
Fig.~8 demonstrates the 
ability of  the synchrotron-proton  model 
to explain the spectra of TeV radiation  of Mkn 501 in both the 
low and high states of the source.  

In the proton-synchrotron model we deal with highly 
magnetized   (100 G or so) condensations of $\gamma$-ray emitting clouds 
of EHE protons, where the magnetic pressure dominates over the pressure of 
relativistic protons.  In the SSC models the situation is exactly opposite. The 
pressure of relativistic electrons is  much larger than the pressure of the B-field
(Aharonian, 2000;  Kino et al., 2001).  Surprisingly,  for TeV blazars 
we do not have an intermediate arrangement !

%  ------------- Fig.8 -------------
\begin{figure}[tb]
\begin{center}
\includegraphics[width=0.68\linewidth]{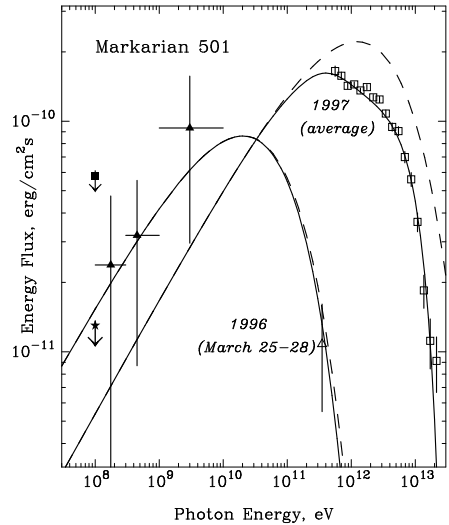}
\caption{Interpretation of TeV radiation of   Mkn 501 in high and low states by
the proton-synchrotron model.
The dashed and solid curves correspond to the spectra 
before and after correction for IG-absorption,  respectively (Aharonian, 2000).}
\end{center}
\end{figure}

Both the leptonic and  hadronic models of TeV blazars  
pose a number of questions.  The answer to some of these  question may 
become  possible  in the near future. 
Such an optimistic view is based on the  recent progress in the 
theoretical (time-dependent) studies of characteristics of nonthermal radiation 
produced in AGN jets,  and, of course, on the unique  data obtained by RXTE and 
ground-based $\gamma$-ray detectors   during the spectacular flares of Mkn 421 
in 2001.  Importantly,  on some occasions  positive signals from Mkn 421 were 
detected also by low-energy air Cherenkov instruments   
STACEE (Ong et al., 2001) and CELESTE  (Le Gallou et al., 2001). 
These results are obtained   at energies around 100 GeV   
at which the IG-absorption could be safely  neglected, and therefore the 
reported fluxes  can be used as  ``calibration''  points when comparing 
the theoretical predictions with experimental data.    

%
%  ------------- Fig.9 -------------
\begin{figure}[tb]
\begin{center}
\includegraphics[width=0.67\linewidth]{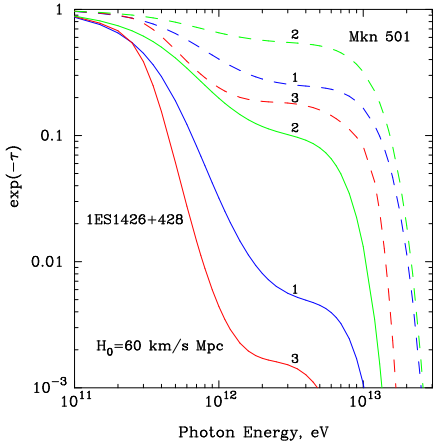}
\caption{Impact of the intergalactic absorption 
on the TeV spectra of  Mkn 501 (upper curves) and 
1ES 1426+428 (lower curves). 
The curves marked as  ``1'', ``2'',  and ``3''  correspond to CIB 
references models  no.1,2, and 3 shown in Fig.~1.}
\end{center}
\end{figure}

Because Mkn 421 and Mkn 501 are located at  almost  {\sl same} 
distances, we can require that any spectral feature attributed to the 
IG-absorption be {\sl exactly the same} for  both sources.  
This is an important circumstance which  can  help to disentangle the 
local features  in the intrinsic spectra from the  contribution induced 
by interactions with  the CIB photons. 
At the same time it is crucial  to have TeV blazars at larger distances, 
so we can exploit the differences in the spectral modification factors. 
Fortunately, in addition to Mkn 421 and Mkn 501,   presently 
we do have  4 more TeV blazar candidates  -- 1ES~2344+5, 1ES~1959+65,
PKS~2155-304 and  1ES~1426+428 -- located at larger distances  (see 
Fig.~2). 

\section{The case of 1ES~1426+428}

Because of the relatively large redshift of 1ES~1426+428
($z=0.129$), the TeV radiation from this extreme BL Lac object
suffers  severe IG-absorption. Therefore  
the  discovery  of TeV $\gamma$-rays  from this source by two  
independent (VERITAS  and HEGRA)  groups
came, to a certain  extent,  as a pleasant surprise.
The spectrum deformation factors
calculated for 3 different CIB models  show that 
above 300 GeV the IG-absorption leads to  significant steepening 
of the  $\gamma$-ray spectrum,   but starting from 1 TeV  to several TeV 
the spectrum is deformed only slightly,  although the suppression
of the absolute flux may be as large  as a factor of  100 (Fig.~9). 
This effect has a simple explanation discussed in  Sec. 3.
%
%  ------------- Fig.10  1ES1426 -------------
\begin{figure}[tb]
\begin{center}
\includegraphics[width=0.68\linewidth]{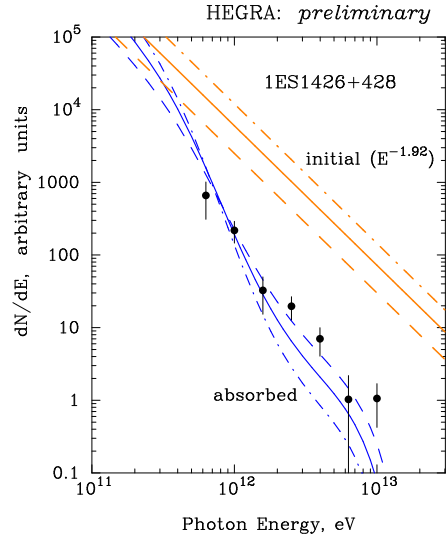}
\caption{The intrinsic  and absorbed $\gamma$-ray spectra 
from 1ES~1426+428 (see the text).}
\end{center}
\end{figure}

Remarkably, the {\em preliminary} spectrum  of 1ES~1426+428
derived from the HEGRA CT system data taken in 1999 and 2000  
(G\"otting et al. , 2001), does show  a steep slope at low  energies 
with a  tendency for a flattening above  1.5  TeV.  
The large statistical errors in flux measurements as well as the 
lack of information  about the intrinsic $\gamma$-ray spectrum
prevent us from definite conclusions. 
Nevertheless,  in Fig.~10 we show the expected spectra of $\gamma$-rays  
{\em assuming} a single power-law intrinsic spectrum with photon index
identical to the  index observed by  {\sl BeppoSAX} in the energy region 
up to 100 keV,  $\Gamma=1.92$   (Costamante et al., 2001).    
The solid curve represents  the absorbed spectrum of $\gamma$-rays
calculated for the CIB model 1 and for the Hubble constant 
$H_0=60 \ \rm km/s / Mpc$.  In order to demonstrate 
the sensitivity  of the intergalactic absorption  effect to the ratio   
of the CIB flux and the Hubble constant, 
$u_{\rm CIB}/H_0$ (for the given  redshift $z$,  and the shape 
of the CIB spectrum,   this  ratio is the only parameter which 
defines the optical depth),  in Fig.~10 we show  
the $\gamma$-ray spectra calculated for  two other values of 
the  ratio $u_{\rm CIB}/H_0$ which are $30 \%$ less (dashed curve) 
and   $30 \%$ more  (dot-dashed curve) compared to  the  the nominal 
value (solid curve).  Hopefully, the future multiwavelength studies
may  allow us to distinguish between these three curves,  thus 
we could be able to  ``measure''  the absolute CIB flux 
at NIR, taking into account that  the current uncertainty of the  
the Hubble constant (presumably) does not  exceed $30 \%$.
 
Independent of details, 
the  pure fact of detection of TeV  $\gamma$-rays from 1ES~1426+428 
is an indicative of a need  for revision of the current 
conceptual view of TeV 
blazars, according to which  the  synchrotron (X-ray) peak in the SED 
dominates over the inverse Compton (TeV)  peak (see e.g. 
Fossati et al. 1998).  Indeed,  
although the {\em detected}  flux of $\gamma$-rays is only 
few times $10^{-12} \, \rm erg/cm^2 s$,  corrected  for IG-absorption
this  flux may  well exceed   
$10^{-10} \, \rm erg/cm^2 s$.  For comparison,  the 
X-ray measurements  (Costamante et al., 2001; T. Takahashi, private communication)  
show fluxes at the level of  $10^{-11} \, \rm erg/cm^2 s$.  
The  TeV luminosity   significantly exceeds 
the X-ray luminosity,  and  apparently 
this should  have an impact on 
the revision of the current TeV-blazar models.        

In particular,   for the 
inverse Compton models   
(for review see Sikora and Madejski, 2001)  
this would require  that the radiation of electrons occurs 
in the regime dominated by Compton losses.  Interestingly, 
this could serve as a good  argument for justification  of  the  above  assumption 
concerning the single power-law {\em intrinsic} TeV spectrum.  Indeed,        
in this regime a  pronounced  feature (spectral flattening) formed in 
the steady-state electron  distribution (caused  by  the structure in the Klein-Nishina 
cross-section)  is largely compensated by the same  Klein-Nishina  effect  
in the radiated  $\gamma$-ray spectrum; thus the $E^{-2}$ type 
{\it acceleration} spectrum of electrons is (almost exactly) transfered  to the 
$\gamma$-ray spectrum (see e.g. Zdziarski and Krolik 1993). Of course, 
in this case we may  expect,  within the one-zone SSC model,  
quite hard synchrotron X-ray spectrum with spectral index $\alpha < 0.5$.
This  actually does not contradict to the   {\sl BeppoSAX}  data
for which the broken power-law model  gives very flat spectrum with 
an index above 10 keV,
$\alpha_2 \approx$0.3-0.4  (Costamante et al., 2001).    

The hard intrinsic TeV spectra of 1ES~1426+428  with constant 
power-law slope and high  $L_\gamma/L_{\rm X} \geq 10$ ratio can be 
explained also by  the {\em proton-synchrotron} model,   assuming 
that  a small,   $\approx 0.1$ fraction of the  proton-synchrotron 
TeV $\gamma$-rays suffer internal absorption. If so, the 
hard X-ray emission could be  attributed to synchrotron radiation of 
the secondary
pair-produced electrons (Aharonian, 2000). The weak point of this model 
is in the requirement of very large, 100 G or so  B-field and 
extremely high acceleration rate of protons.

%---------------------------------- ------------- Fig.11 -------------
\begin{figure}[t]
\begin{center}
\includegraphics[width=0.75\linewidth]{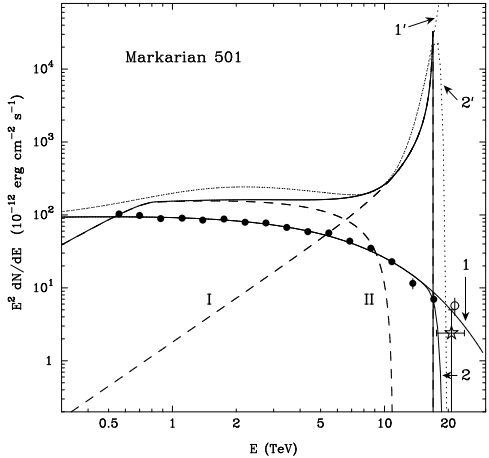}
\caption{Inverse Compton spectrum of cold unshocked 
ultrarelativistic jet with a  Lorentz factor of  bulk motion
$\Gamma = 3.33 \times 10^7$. 
The radiation component associated with Comptonization of 
ambient optical photons is shown by
dashed curve I. 
The dashed curve  II is the  
residual of the total TeV source emission 
(after subtraction of the unshocked wind component I),  and 
formally   could be attributed  either to the Comptonization of the bulk 
motion on ambient  FIR photons or  to the radiation 
of blobs in shocked jet.   
The heavy solid line represents the 
superposition of these two components.
Fits to the observed flux of Mkn~501 are shown by thin solid lines.
Curve  1 corresponds to the fit given by equation (b) in Fig.~3,  
and curve  2 -- to the steepest possible spectrum 
above 17 TeV based on the recent reanalysis of Mkn~501 HEGRA data 
(Aharonian et al., 2001a).  The intrinsic (absorption-corrected) spectra of Mkn~501,
corresponding to the  fits to the observed points  ``1'' and ``2'', 
are shown by dotted lines ``$1^\prime$'' and ``$2^\prime$'', 
respectively  (from Aharonian et al., 2002).}
\end{center}
\end{figure}
\section{``IR background - TeV gamma-ray crisis'' ?}

 % ------------------------------ FIG 12-----------------
%
\begin{figure*}[t]
\includegraphics[width=14.0cm]{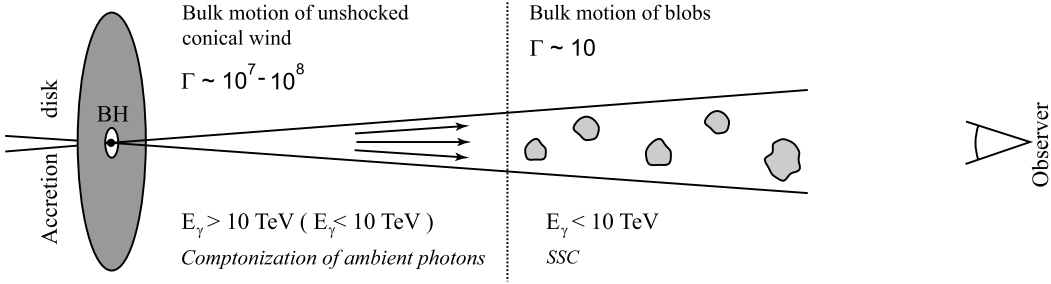}
\caption{ Sketch of the two-stage $\gamma$-ray production scenario 
in Mkn~501. At the first stage,  the highest energy
$\gamma$-rays above 10 TeV with a 
sharp spectral form  (``pile-up'') 
are produced due to Comptonization of the ambient 
optical photons  
by the cold  ultrarelativistic jet with bulk motion Lorentz factor 
$\Gamma \sim 10^7-10^8$. Low energy $\gamma$-rays, 
$E_\gamma  \leq 10 \ \rm TeV$ can be also produced due to  Comptonization 
of ambient FIR photons with broad spectral distribution. 
At the second stage,   relatively low energy ($E_\gamma  \leq 10 \ \rm TeV$)
$\gamma$-rays are produced  in the blobs moving with Lorentz factor 
of about $\Gamma \sim 10$ in accordance with 
the standard   SSC model (from Aharonian et al., 2002).}
\end{figure*}
From the discussion of the previous section we may  summarize  that 
the basic temporal and spectral characteristics of TeV blazars can be described,
at least qualitatively, by the  the current leptonic, and perhaps 
also  hadronic models.  At the same time  all these models 
would fail to explain the sharp pile-up  which  may appear at the end of 
the ``reconstructed'' spectrum of Mkn 501, if the reported FIR fluxes 
correctly describe the level of  CIB (see Fig.~3).   
Motivated by such a non-standard spectral shape, 
recently  several extreme hypotheses have been proposed to overcome the 
``IR background - TeV gamma-ray crisis'' (Protheroe and Meyer 2000).   
In particular,  Harwit et al. (2000) 
suggested  an interesting   idea that the HEGRA highest energy events 
are due to Bose-Einstein condensations interacting  with the air 
atmosphere.  
%However,   subsequently the HEGRA collaboration has demonstrated 
%(Aharonian et al. 2000)  that  the detected shower characteristics are in fact in good 
%agreement with the predictions  for the   
%events initiated by {\it ordinary}  $\gamma$-rays.
Another, even more dramatic hypothesis
-- violation of the  Lorentz invariance -- has been 
proposed   to solve this problem  (e.g. Kifune, 1999;  Stecker and  
Glashow, 2001).  Two more proposals   could be added    to the list of  
``exotic''  solutions:   
(i)   assuming that Mkn 501 is located at a distance $\ll 100$~Mpc; 
or (ii) assuming that TeV $\gamma$-rays  from  
Mkn~501 are not direct representatives of  primary radiation of the source, 
but  are formed  during the development 
of  high energy electron-photon cascades in the intergalactic medium
(Aharonian et al. 2001b).
At first glance, both ideas  seem relatively  ``neutral''.   But  in fact they  
do contain  dramatic  assumptions.   
While  first hypothesis implies  non-cosmological 
origin of the redshift  of Mkn 501,  the second  hypothesis requires 
extremely low intergalactic magnetic field of order of 
$10^{-18}$ G or less.   

Although very fascinating,  the appeal for  
revisions of essentials of modern physics and astrophysics
seems to be too premature. 
The nature of the FIR  isotropic emission detected by COBE 
is  not yet firmly established, and  it is quite possible that the bulk of the 
reported flux, especially below 100 $\mu \rm m$, is a result of superposition of
different local backgrounds.  It should be noticed, however,  that not only 
the flux at $60 \mu \rm m$,  but also
the CIB  fluxes at longer,  $\lambda \geq 100 \mu \rm m$  wavelengths
are responsible  (albeit in a less distinct form)   
for the   appearance of the pile-up, unless we assume very specific CIB 
spectrum in the  MIR-to-FIR transition region 
(Aharonian et al., 1999a; Coppi and  Aharonian, 1999b;  
Guy et al., 2000; Renault et al., 2001).  
Therefore it is  important  to explore other possibilities for formation  
of  pile-ups in the spectra of TeV blazars. 
Motivated by this,  recently we have proposed  a new, 
{\em non-acceleration} scenario which postulates 
that the $\geq 10 \ \rm TeV$ 
radiation of  Mkn~501 is a result of the  {\em bulk motion
Comptonization} of ambient  low-frequency photons  
by  a cold ultrarelativistic conical wind  (Aharonian et al., 2002). 

The  relativistically moving plasma outflows  in forms of    
jets or winds,  are  common for many  astrophysical phenomena  
on both galactic  or extragalactic scales. 
Independent of  the origin of these  relativistic outflows, the concept of 
the jet  seems to be the only successful approach to understand  the complex 
features  of  nonthermal radiation of {\em Blazars,  Microquasars}  and 
{\em GRBs}. The Lorentz factor of such outflows could be extremely 
large. In particular in the  Crab Nebula the Lorentz-factor of the 
MHD wind  is estimated   between  $10^6$ and $10^7$  (Rees and  Gun, 1974;
Kennel and Coroniti, 1984).   Meszaros and Rees (1997) have
shown that  in the context of cosmological GRBs  
the magnetically dominated jet-like  outflows from stellar mass 
black holes may attain extreme  Lorentz factors exceeding  $10^6$. 
The  conventional  Lorentz factors of jets  in 
the inverse Compton models of $\gamma$-ray blazars,
are  rather  modest,  $\Gamma \sim 10$. However  there are no apparent 
theoretical  or observation arguments against the bulk motion 
with much larger  Lorentz-factors  (see e.g. discussion by  Celotti et al., 1998).
Because of  existence of dense photons fields  in the 
inner sub-parsec region of sources like Mkn 421 and Mkn 501,  
the Compton optical depth $\tau_{\rm C}$ could be as large as 1. 
Obviously, for our model 
the most favorable  value for $\tau_{\rm C}$   
lies   between 0.1 (in order to avoid  huge  energy requirements to the 
outflow) and 1 (in order to avoid the Compton drag).
In particular,  we have to assume that in the proximity of the black hole 
the outflow should be Poynting flux dominated, and  only at 
large  distances from the central object,  
where the photon density is significantly  reduced,   
the major part of the  electromagnetic energy  is transfered to 
the kinetic energy of bulk motion. 

Due to the extremely large  
Lorentz factors exceeding  $10^7$, the Compton scattering on the 
ambient NIR/optical photons with energy more than 1 eV  proceeds in deep 
Klein-Nishina regime; therefore the $\gamma$-radiation  should have
a   very narrow distribution with energy  
$E \approx E_{\rm e}= m_{\rm e} c^2 \Gamma$.  
Meanwhile,   the IC scattering  on ambient far IR photons surrounding 
the central source,  still takes place in the 
Thomson regime,  and thus results  in  a smooth broad-band  spectrum. 

Fig.~11 demonstrates  that the overall {\em absorption-corrected} 
(for the CIB at FIR close to the  extreme CIB model  no.3)  spectrum  of 
Mkn~501 can be satisfactorily explained  in  the terms of the  
inverse Compton emission of the cold jet with bulk Lorentz factor 
$\Gamma=3.33 \times 10^7$,   assuming an
ambient radiation field with a  narrow  (Planckian)  type 
radiation with temperature  $kT= 2 \, \rm eV$ (dashed curve I).
The dashed curve  II  which formally is  the   
residual from the subtraction of  the unshocked wind component radiation 
from  the intrinsic (reconstructed) TeV emission (solid heavy line), 
can be attributed  to the bulk Comptonization on ambient    
far IR photons produced e.g. by cold clouds surrounding the source. 
Alternatively,  the  ``residual component''  (curve II)  
can be referred to the SSC (or any other) radiation  component 
of blobs in shocked jet.  This two-stage ({\it pre-shock} 
plus {\it post shock}) 
scenario of formation 
of TeV $\gamma$-ray emission is schematically 
illustrated in  Fig.~12 

The possibility to disentangle  the multi-TeV emission with a characteristic 
sharp pile-up at the very end of the spectrum, 
$E_\gamma \simeq m_{\rm e} c^2 \Gamma$, from the sub-10~TeV emission 
associated with the shocked structures (e.g. blobs) in the jet,   
not only solves the  possible ``IR background -- TeV gamma-ray crisis''   but   also 
allows more relaxed parameter space for  interpretation 
of X-rays and the  {\it remaining} ``low''  energy ($\leq 10 \ \rm TeV$)
$\gamma$-rays   within the conventional  SSC scenario. 
Consequently,  this  offers  more
options for interpretation of X-ray/gamma-ray correlations.  
If the overall TeV radiation of Mkn~501 indeed consists 
of two, {\it unshocked} and {\it shocked}  jet  emission  components, 
we may expect essentially different time  
behaviors of these components. In particular, the ``unshocked jet'' 
($\geq 10 \, \rm TeV$)  component  should arrive earlier than the 
SSC components  consisting of synchrotron X-  and 
sub-10 TeV IC $\gamma$-rays.  Therefore, 
an important test of the suggested two stage 
scenario of the TeV radiation of jets 
would be  the search  for correlations (or lack of such 
correlations) of $\geq 10 \ \rm TeV$  radiation 
with both the low energy (e.g. 1-3 TeV)  $\gamma$-rays and synchrotron 
X-rays.  The low statistics of  (heavily absorbed) $\gamma$-rays above 10~TeV 
makes the search for such correlations rather difficult, and requires 
ground-based instruments with huge, $\gg 0.1 \, \rm km^2$ 
detection areas in this energy domain.  The new generation IACT 
 arrays like CANGAROO-3, H.E.S.S. and VERITAS should be able
to perform  such correlation studies.   

It should be noticed, however,  that the study of the signatures of 
IG-absorption of $\gamma$-rays from  Mkn 421 and Mkn 501
does not contain  adequate  information  about  CIB 
at $\lambda \geq 40 \mu \rm m$. This information  is 
essentially  lost due to the heavily  absorbed 
$\gamma$-rays above 20 TeV.    
This objective  is rather contingent on discovery of nearby 
extragalactic TeV sources within $\approx$ 50 Mpc,   
dictated  by  the  condition that the mean free path of $\gamma$-rays  
at these energies should  not  significantly exceed the distance to the 
source (see Fig.~2).  Despite the lack of  very close  blazars, 
other potential extragalactic $\gamma$-ray sources  like the nearby 
radiogalaxies Centaurus A and M~87, and perhaps also  the startburst galaxies 
Arp 220,  M~82 and  NGC~253  may   (hopefully) provide us with   
multi-TeV  $\gamma$-rays  for such   important studies.  

%
%  ------------- Fig.12 halos -------------
\begin{figure}[tb]
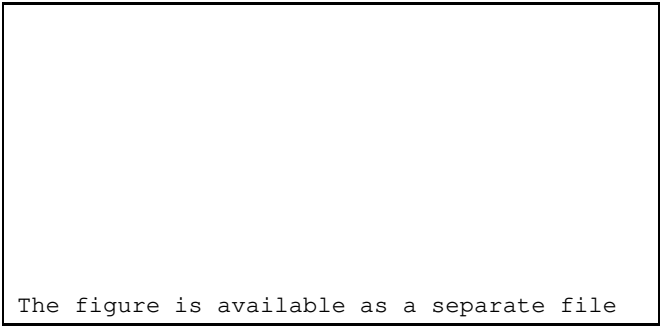

\begin{center}
\fbox{\parbox{\linewidth}{\rule{0pt}{4cm}
\texttt{The figure is available as a separate file}
}}
\caption{Illustration of  the process of  formation of Pair Halos.}
\end{center}
\end{figure}
Yet, the  distant TeV  sources may offer an additional, 
and perhaps even more informative channel for derivation  of CIB fluxes 
at FIR wavelengths. Such a possibility is based on the study of angular and spectral 
characteristics of hypothetical giant {\em Pair Halos} expected around powerful 
nonthermal extragalactic sources  (Aharonian et al., 1994).       

\section{Pair Halos and CIB}

Because of IG-absorption, the bulk of primary TeV $\gamma$-rays   
from distant extragalactic sources is not visible.  The energy of these photons   
however is not lost --  the electromagnetic cascade initiated by the absorbed 
$\gamma$-rays  can be observed~!  The cascade in the IG  
medium (e.g.  Berezinsky and Kudriavtsev, 1990; Protheroe and Stanev, 1993;
Coppi and Aharonian, 1997)   
is supported  by 2 processes: (i) by inverse Compton scattering on the photons of 
2.7 K MBR and (ii) by  photon-photon pair production  on CIB 
(for energies above 100 TeV - also on 2.7 K MBR).  When the magnetic field 
near the source  is sufficiently large, $B \geq  10^{-9}$ G 
(generally on the scales of  tens of  Mpc, but if the spectrum of 
primary $\gamma$-rays extends  into PeV region - within tens of kpc), 
i.e. when the mean free path of cascade  electrons 
(at least at the initial stages) is large compared to Larmor radii, 
the cascade will generate,  even for a highly beamed primary source, an 
extended isotropic Pair Halo (see Fig.~13).  While the formation of Pair Halos 
is unavoidable for  {\em all} extragalactic  objects 
with spectra extending beyond 10 GeV,  only relatively 
compact  Pair Halos formed  around (multi) TeV sources could be detectable.  
The superposition  of the cascade radiation  of  all   
Pair Halos formed  in the Universe  
may contribute significantly   to the diffuse 
GeV $\gamma$-ray background  (Coppi and Aharonian, 1997). 

The halo radiation can be distinguished  by its characteristic variation in 
spectrum  and intensity with angular distance  from the central source,   
which weakly depends on the details of the source model,   in 
particular  on the energy  spectrum and the   geometry  of the primary 
source,  especially  when  a hard primary  $\gamma$-ray spectrum 
extends to  PeV energies.

The energy $E_\gamma$ and the arrival angle $\theta$  of the 
detected photon  from a source at a distance $d$ 
are  determined, on average, 
by the energy of the parent electron 
$E_{\rm e}  \approx (E_\gamma/4 k T_0)^{1/2} m_{\rm e} c^2$ ($T_0=2.7 \ \rm K$), 
and  the  distance $l$ from 
the central source where this electron is produced. 
Because the mean free path of $\gamma$-rays $\Lambda$ decreases with energy 
(see Fig.~2),  the energy and site of production of the secondary electron
are defined, on average, by the ``last $\gamma$-CIB interaction'',
$E_{\rm e} \approx    E_{\gamma,0}/2$,  therefore the energy $E_\gamma$ 
and arrival angle $\theta$  of the {\em detected}  $\gamma$-rays contain information
about  their ``grandparents'' - $\gamma$-rays with energy 
$E_{\gamma,0} \simeq 34 (E_\gamma/1 \ \  \rm TeV)^{1/2} \ \rm TeV$, 
and mean free path   $\Lambda(E_{\gamma,0}) = d \cdot \theta$.
Correspondingly,  the {\em measured} 
mean free path   $\Lambda(E_{\gamma,0})$  
should tell us about the density of 
CIB at  $\lambda \sim  50 (E_\gamma/1 \ \rm TeV)^{1/2} \ \mu \rm m$. 
Thus,   detections of Pair Halos  surrounding Mkn 421 and Mkn 501 
at  1 TeV should provide information on  CIB 
around 50  $\mu \rm m$, provided that the primary spectrum extends beyond 30 TeV. 

Recently the HEGRA collaboration made the  first
attempt  to probe the existence of  a Pair Halo in direction of  
Mkn~501,  using the data base accumulated during the 
period of observations from 1997 to 1999 
(Aharonian et al., 2001c).  The derived limit on the halo flux  
$J_{\rm PH} (\geq 1 \, \rm TeV) \sim  10^{-8} \rm ph/cm^2 s \ sr$
within $1^{\circ}$  is unfortunately  above the  theoretical  
expectations  based on  the energy budget constraints. Indeed,   
Mkn 501 actually is not an  ideal  candidate for a halo source given 
its  relatively modest luminosity  and the fact that the preferable 
distance  to observe halo is in the range of several 100 Mpc to 
1 Gpc;  consequently,  the most effective searches 
can be performed  by detectors with energy threshold 
around 100 GeV (Aharonian et al., 1995).  
In this regard, the BL Lac objects   
1ES~1426+428  at $z=0.129$ and PKS~2155-30 at 
$z=0.116$    with $\gamma$-ray luminosities exceeding the ``quiescent'' 
TeV luminosity of Mkn 501 al least by  a factor of $10$,   seem to be  
more  suitable objects  for producing  {\em detectable} 
Pair Halos. The forthcoming   H.E.S.S. array of 10m IACTs 
equipped with  optimal for this purpose  $5^{\circ}$ FoV high resolution cameras,  
will  be the first  instrument  for a sensitive 
search for  Pair Halos  at $z \sim 0.1-0.2$. 

Generally, the blazars are  are relatively weak sources -- 
they seem very bright  because of   the Doppler boosting   (proportional to 
$\delta_{\rm j}^4 \geq 10^3$).  The distant powerful radiogalaxies 
and quasars with large-scale jets (as potential accelerators of ultrahigh energy 
cosmic rays) may  appear more promising sites for production  
of detectable -- {\em strong} and {\em compact} --  Pair Halos  (Aharonian, 2002).  
The detection and identification 
of such halos from cosmologically  distant  objects beyond  $z = 1$  
is possible at energies of $\gamma$-rays  of about  10 GeV. 
While the spectral cutoffs in the energy spectra of 
the direct and halo emission components  are caused   by 
absorption in the diffuse  UV background radiation,  the 
angular size  of the halo  contains   information 
about the CIB  at  several to 10 $\mu \rm m$ wavelengths 
at the epoch corresponding to the redshift of the central 
source $z$. This is a unique (not available by other means) 
channel which provides us with model-independent 
information about the cosmological  evolution of CIB.
Such studies can be effectively 
performed  by GLAST (e.g. Gehrels and Michelson 1999) 
and by future sub-10 GeV threshold high-altitude 
IACT arrays like 5@5 (Aharonian et al. 2001).   

\section{Summary}
The  energy-dependent mean free path of $\gamma$-rays 
in the intergalactic medium    at TeV energies 
does not  exceed several 100 Mpc. 
Therefore   VHE $\gamma$-rays   from  TeV blazars  
arrive with  significantly distorted spectra. Our  limited  knowledge 
of  CIB   results in  large  uncertainties
in the  reconstructed (corrected for intergalactic absorption) intrinsic  
$\gamma$-ray spectra.  Consequently, in spite of  good quality of 
spectral measurements  of two strongest TeV blazars, Mkn 421 and Mkn 501,  
nowadays  we are faced  with  a challenge   - our 
understanding of  radiation  processes in relativistic  jets is neither 
complete nor conclusive.   Nevertheless,  the gamma-ray astronomers believe 
that  eventually  (soon ?) they  will learn  to identify confidently the radiation 
mechanisms, to fix/constrain  the relevant model parameter space, and 
calculate robustly   the intrinsic $\gamma$-ray spectra based on 
the multiwavelength studies of spectral and 
temporal characteristics of  blazars obtained simultaneously at X-ray and TeV 
bands on  sub-hour timescales.
Then it will be possible  to estimate unambiguously   the effect of  
the intergalactic $\gamma$-ray absorption,  and thus to  
infer robust  information about the CIB fluxes.  
Moreover, the studies of angular  and spectral 
properties of giant Pair  
Halos   formed  around  powerful  extragalactic multi-TeV sources  
may  provide us with a unique tool 
for  derivation of   the spectra and absolute fluxes  of CIB at 
{\sl different cosmological  epochs z}, and thus to probe 
the evolution of galaxies in past.

\end{document}